\documentclass{aa}
\usepackage{graphics}
 
\begin{document}

\thesaurus{11(13.18.1; 13.25.2; 11.02.2; 11.10.1)} 

\title{Multiwavelength observations of 26W20, a radio galaxy which
displays BL Lac characteristics.}

\author{J. D.Silverman\inst{1}, D. E. Harris\inst{1}
        \and W. Junor\inst{2}}

\institute{Harvard-Smithsonian Center for Astrophysics, 60 Garden
Street, Cambridge, MA 02138
	\and Department of Physics and Astronomy, University of New Mexico, 800 Yale Blvd., NE, Albuquerque, NM 87131   }
 
\date{Received date / Accepted date}
\offprints{D.E. Harris}

\titlerunning{26W20, a radio galaxy which displays BL Lac characteristics}
\authorrunning{Silverman et al.}
\maketitle

\begin{abstract}

\object{26W20}, a radio galaxy located at a projected distance of 2.2 Mpc from
the center of Abell 754 (z=0.054) exhibits core emission properties
similar to those of a BL Lac object.  New radio observations and VLA archival
data show parsec scale features in the core (\(L_{r}\sim10^{41}\)
ergs s$^{-1}$) with a flat spectrum.  From observations with ROSAT, we
demonstrate that the X-ray core emission \((L_{x}=3\times10^{43}\)
ergs s$^{-1}$) is non-thermal and variable.  An 18\% increase in the X-ray
luminosity was observed in two observations separated by five days.
The optical morphology is that of a large, normal elliptical
galaxy (\(m_{v}=15.6)\) and the optical spectrum lacks strong
emission lines.  If the innermost segment of the radio jet is close
to our line of sight, Doppler boosting could explain both the observed
X-ray intensity and the absence of emitting line regions.

\end{abstract}

\section{Introduction}

26W20 was discovered in the Westerbork radio survey by Harris {\it et
al.} (1980).  The galaxy lies in a sub-cluster of \object{Abell 754}.  The
structure of the cluster and the reality of the sub-cluster containing
26W20 are detailed by Kriesslar and Beers (1997).

The radio morphology and classification has been described as a
Fanaroff and Riley class I object, either a 'tailed radio galaxy'
(TRG) or a single-sided lobe (Harris {\it et al.} 1980; Harris,
Costain and Dewdney, 1984 - 'HCD' hereafter).  A single jet/tail is
seen extending 170 kpc from the core (\(H_{o}\)= 50 km s$^{-1}$
Mpc$^{-1}$) and ending in a low surface brightness lobe (Fig. 1).
Because no jet or lobe was detected on the other side of the galaxy,
and since the host galaxy is a member of a small group of galaxies, a
TRG model was favored over a single-sided jet model (HCD).

Galaxies with X-ray bright nuclei are usually characterized by strong
emission lines.  26W20 is one of four known galaxies with a high
apparent X-ray luminosity from the nucleus but with only weak or
undetectable emission lines.  The other three are \object{3C264} (Baum
{\it et al.}  1997), \object{J2310$-$43} (Tananbaum {\it et al.}
1997) and \object{PKS 2316$-$423} (Crawford and Fabian 1994).  Nuclear
parameters are given in Table 1.

26W20 was first observed in X-rays with the EINSTEIN Observatory's
imaging proportional counter (IPC) and high resolution imager (HRI).
The X-ray source was unresolved with the HRI (FWHM=5\arcsec) and the
IPC spectrum could be equally well fit by a power law with an energy
index of 0.8 \(\pm\)0.4 or a thermal bremsstrahlung spectrum with kT=3
keV (constrained to be greater than 1 keV).

Motivated to understand the head-tail morphology of 26W20, we have
obtained new radio and X-ray measurements to study the inner jet
region.  While there is possible evidence for a TRG model, these
observations reveal core emission properties similar to those of a BL
Lac object. However, there is no evidence for excess optical emission
from the nucleus.  This distinguishes 26w20 from a typical BL Lac object.

We use ROSAT HRI data to provide upper limits on hot ambient gas which
would be required to provide the 'ram pressure' if nuclear gas is to
be 'stripped' from the galaxy. MERLIN observations were obtained to
see if twin jets are present in the core and to see if they show
severe bending necessary for a beaming model in order to change the
jet from being close to the line of sight to that observed for the
large scale tail which lies more in the plane of the sky.

\scriptsize
\begin{flushleft}
\begin{tabular*}{90mm}{lllll} 
\multicolumn{5}{c}{Table 1.} \\
\multicolumn{5}{c}{Comparative parameters for the nucleus of similar sources} \\
\hline\\
Parameter&26W20&J2310-43$^{1}$&3C 264$^{2}$&PKS 2316-423$^{3}$\\
\hline
Redshift..............&0.054&0.0886&0.0215&0.0549\\
l$_{r}$(10$^{30}$ cgs)$^{a}$&7.8&21.2&5.2&70.3\\
L$_{x}$($10^{43}$ergs s$^{-1}$).&3.7$^{b}$&14.5$^{c}$&0.88$^{c}$&8.6$^{c}$\\
M$_{V}^{d}$.....................&$>$-19.6&$>$-20.1&$>$-19.8&$>$-20.8$^{e}$\\
$\alpha_{r}$......................&0.25&$\sim$0&$\sim$0&0.67\\
$\alpha_{x}$......................&1.3&1.4&1.3&---\\
Emission Lines...&none&none&weak$^{f}$&weak\\
\label{sources}
\end{tabular*}
\end{flushleft}	
References: (1) as reported in Tananbaum et al. 1997 (2) as
reported in Tananbaum et al. 1997, Elvis et al. 1981; (3) Crawford and
Fabian 1994\\
\\
Notes: (a) Monochromatic luminosity at 5 GHz in units of ergs s$^{-1}$
Hz$^{-1}$; (b) Energy range of
(0.5-3.0 keV); (c) Energy range of (0.1-2.4 keV); (d) The core
magnitude is assumed to be less than 1/10 of the magnitude of the
whole galaxy; (e) This lower limit is due to the
inclusion of optical emission not associated with the nucleus; (f)
An extended region of 400 km s$^{-1}$ wide
H$\alpha$+[NII] emission is seen (Baum {\it et al.}  1997); 
\normalsize

\section{Observations and Data Reduction}
\subsection{Radio}
\subsubsection{Low Resolution}

\begin{figure}
  \resizebox{\hsize}{!}{\includegraphics{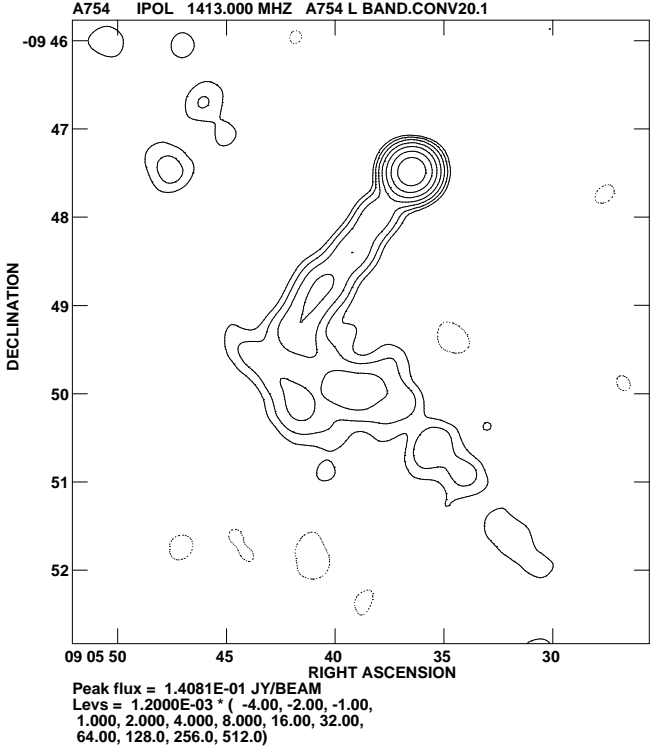}}
  \caption{A naturally weighted $\lambda$21 cm map of 26W20 from
1980 VLA observations.  The image has been restored with a $20\arcsec$ 
circular beam in order to delineate the low surface brightness features.}
  \label{20cm}
\end{figure}

The archival 21 cm VLA radio data of 26W20 have been reprocessed
(Figure \ref{20cm}).  This map was made with a hybrid configuration of
antennas and a 30k$\lambda$ taper.  At 1413 MHz, the core has a flux
density of 165 mJy; the tail, 87 mJy; and the lobe, 171 mJy. The total
flux density of the source is 435$\pm$40 mJy.  The TRG model requires
the galaxy to be moving with a sizeable fraction of its velocity
vector perpendicular to our line of sight.  In this model, the twin
jets point initially parallel and anti-parallel with the
line-of-sight.  They are then bent back over short spatial scales in
such a way that the twin tails appear superimposed.

\subsubsection{High Resolution}

\begin{figure}
  \resizebox{\hsize}{!}{\includegraphics{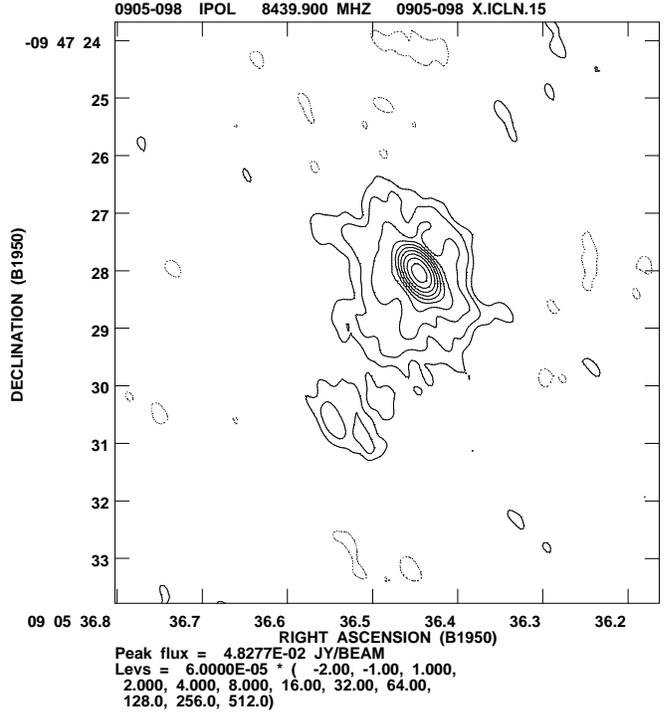}}
   \caption{VLA tapered $\lambda3.6\,$cm image of the core of 26W20.
The restoring beam is 0.44\arcsec $\times$ $0.25\arcsec$ in PA $= 33.3^{\circ}$.}
  \label{4cm}
\end{figure}

Imaging of the VLA archival data (Figure \ref{4cm}) at 3.6 cm shows an
unresolved core with a peak flux density of 48 mJy/beam.  Using the FWHM as
the angular extent, the scale size of the core is $\sim1.5$ kpc.

\begin{figure}
  \resizebox{\hsize}{!}{\includegraphics{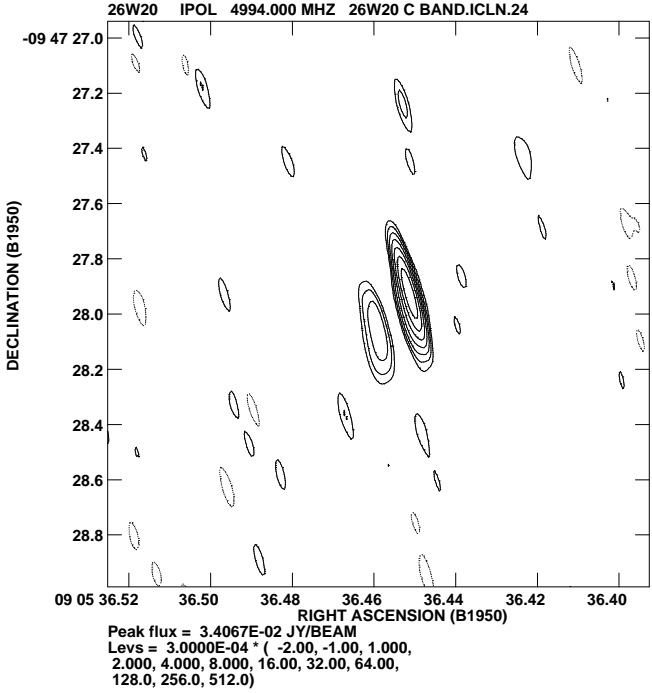}}
   \caption{MERLIN $\lambda6\,$cm image of the core of 26W20.  The
synthesized beam is 0.20\arcsec $\times$ 0.03$\arcsec$ in PA $=16.0^{\circ}$.}
\label{6cm}
\end{figure}

We recently mapped 26W20 with the MERLIN array at 6 cm (Figure
\ref{6cm}).  With an integrated flux density of 45 mJy, 70\% of the
emission measured by the VLA is contained within a compact region of
scale size $\sim$100 pc.  From the image it is seen that we were able
to detect a second component at nearly the same position angle as the
jet/tail in the VLA maps.
	
\subsubsection{Radio Spectrum}

\begin{figure}
  \resizebox{\hsize}{!}{\includegraphics{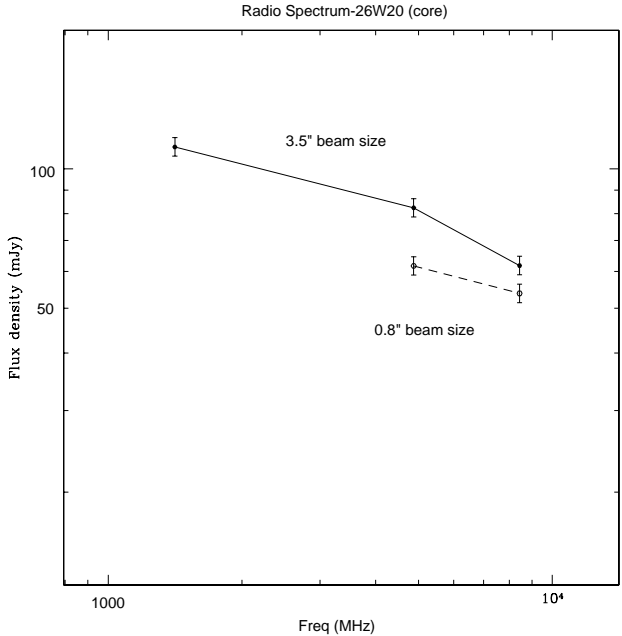}}
  \caption{Radio spectrum using different beam sizes.}
  \label{radio_spectrum}
\end{figure}

Table 2 lists the flux densities for all the observations.  Figure
\ref{radio_spectrum} shows the radio spectrum of the core using two
beam sizes of $0.8\arcsec$ and $3.5\arcsec$.  The spectral index,
\(\alpha (S_{\nu} \propto \nu^{-\alpha})\), is 0.25$\pm0.10$ for the
$0.8\arcsec$ beam.

\scriptsize
\begin{flushleft}
\begin{tabular}{lllll}                              
\multicolumn{5}{c}{Table 2.} \\
\multicolumn{5}{c}{Flux densities for the core of 26W20} \\
\\
\bf Freq&\bf MJD&\bf Beamsize&\multicolumn{2}{c}{\bf Flux Density(S)} \\
(MHz)& &(arcsec)& Peak(mJy)& Integrated(mJy) \\ \hline
1413$^{\it 1}$&44360&3.5&111.6(5.6)&---\\  
4873$^{\it 1}$&44360&3.5&82.4(4.1) &--- \\
4873$^{\it 2}$&44360&0.8&61.7(3.1)&--- \\
4994$^{\it 3}$&50417&0.18$\times$0.03&34.1(1.7)&45(2.2)\\
8439$^{\it 4}$&48982&0.4$\times$0.25&48.3(2.4)&61.8(3.1) \\
8439$^{\it 5}$&48982&0.4$\times$0.25&---&53.8(3.0) \\
10550$^{\it 6}$&&69&75(2.8)&--- \\
\label{radio_flux}
\end{tabular}
\end{flushleft}
Notes:
\\
The uncertainties are given in parenthesis. \\
1 Obs-date:1980Jun01; 50 k$\lambda$ taper \\
2 Obs-date:1980Jun01; HCD, fig 6 \\
3 Obs-date:1996Dec; MERLIN \\
4 Obs-date:1992Dec26; Integrated flux using a $3\arcsec\times3.5\arcsec$ box. \\
5 Obs-date:1992Dec26; Integrated flux using a 1$\arcsec\times1\arcsec$ box. \\
6 Obs-date:91/92; MPIfR, Mack {\it et al.} (1993).  In that
compilation of 10 TRGs, a table is given of previous observations; but most or all appear to use large beams and 
  so do not provide useful data for our purposes.
\normalsize
\subsection{X-ray}

Due to the close proximity of 26W20 to the cluster Abell 754, there
are several observations with ROSAT and EINSTEIN for which 26W20 is in
the field of view.  The galaxy was first detected with the EINSTEIN
observatory and imaged with the IPC and HRI
\((L_{x}=2.2\times10^{43}\) ergs s$^{-1}$; HCD).  Five ROSAT observations
using the HRI and the Position Sensitive Proportional Counter (PSPC)
are available spanning two years (Table 3; three off-axis PSPC, one
on-axis PSPC and one on-axis HRI).  All luminosity measurements are
calculated for the energy range 0.5 to 3.0 keV with one sigma errors
associated with the count statistics.  Data analysis was performed
with the PROS package under IRAF.

\begin{flushleft}
\scriptsize
\begin{tabular}{lllll}                              
\multicolumn{5}{c}{Table 3.} \\
\multicolumn{5}{c}{X-ray Source Parameters} \\
\\
\bf Instr.&\bf Obs&\bf Exp. time&\bf Cnt rate&\bf
 Luminosity \\
 &\bf date &(sec)& (cts s$^{-1}$) & \((10^{43}\) ergs s$^{-1}$) \\ \hline
EO&1979/80& & &2.2(0.65) \\
PSPC$^{2}$&21/11/91&2322&0.154&3.88(0.23) \\
PSPC$^{1}$&05/11/92&13499&0.153&3.12(0.07) \\
PSPC$^{2}$&10/11/92&6359&0.146&3.67(0.15) \\
HRI$^{1}$&22/11/92&24225&0.066&3.70(0.10) \\
PSPC$^{2}$&06/11/93&8160&0.112&2.85(0.10) \\
\label{xray_table}
\end{tabular}
\end{flushleft}
\scriptsize
Notes:\\
(1) On-axis observations. \\
(2) Off-axis observations. The count rates have not been corrected for
vignetting. \\
\normalsize

\subsubsection{Spatial Extent}

\begin{figure}
  \resizebox{\hsize}{!}{\includegraphics{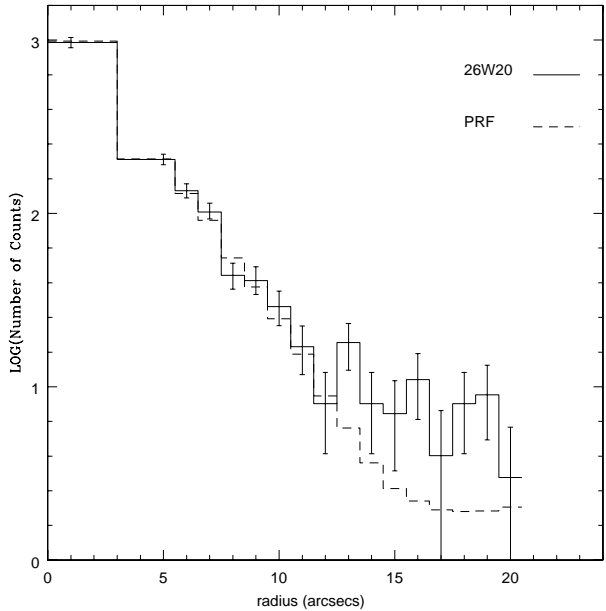}}
  \caption{A radial profile of the ROSAT HRI data.  Counts are binned in $1\arcsec$ intervals.}
  \label{radial}
\end{figure}

The X-ray core emission is unresolved with the ROSAT HRI.  The
radial profile (Figure \ref{radial}) shows the close fit of the count
distribution to the ROSAT point response function.

There is no conclusive evidence for the presence of any significant
extended emission in the local environment of 26W20.  The low level
emission seen in the $12\arcsec$ to $20\arcsec$ region at about 1\% of
the peak intensity can be attributed to instrumental effects.  As
shown in Figure 1 of The HRI Calibration Report (David {\it et al.}
1995), there is a discrepancy in the fit to the point response
function in the $10\arcsec$ to $20\arcsec$ range as determined using a
long \object{HZ43} observation.

\subsubsection{Spectral Fit}

Analysis of archival X-ray data from the ROSAT PSPC and our HRI data
(including EINSTEIN) is presented in Figure \ref{var} and Table 3.
The energy index and the galactic column density were allowed as free
parameters.  The values of the fitted column density agreed closely
with the galactic value of $4.58\times10^{20}$ cm$^{-2}$.  Using the
long, on-axis PSPC observation (13 ksec), we find for a power law
model fit with 29 degrees of freedom, $\alpha=1.32\pm0.17$, log
\(N_{H}\)=20.90$\pm$0.04 and $\chi^{2}=8.30$.  For a Raymond-Smith
thermal model, the fitted spectral parameters are kT=$4.3\pm0.7$ keV;
log $N_{H}=20.59\pm0.03$, and $\chi^{2}=67.54$.  On the basis of the
goodness of fit, the core emission is described more favorably by a
non-thermal model.
	
The best fit spectral parameters for each individual observation are
shown in Figure \ref{var}.  The slight increase in the energy index is
within the errors.

\subsubsection{Variability}

\begin{figure}
  \resizebox{\hsize}{!}{\includegraphics{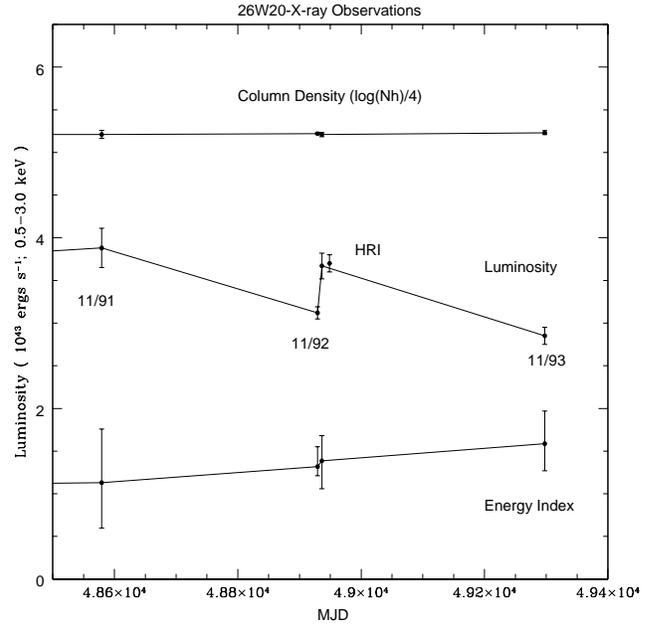}}
\caption{X-ray variability and spectral parameters of 26W20 with one
sigma errors.  The curves connect to the average EINSTEIN values of 1979/1980 
(off the left side of the graph): L$_{x}$(EO)=2.2$\times10^{43}$ergs s$^{-1}$,
$\alpha$=0.8, log N$_{H}$=20.68.}
\label{var}
\end{figure}

All ROSAT luminosities (Figure \ref{var}; Table 3) have been
calculated with a power law model with an energy index of 1.32 and a
log of the galactic $N_{H}$ of 20.90.  These values are the best fit
parameters from the longest on-axis PSPC observation.  A decrease of
the luminosity from PSPC measurements is relatively mild over two
years from 1991 to 1993. However, an 18\% increase in the luminosity
occurred within five days between two PSPC observations in 1992.  An
HRI image about two weeks later confirms this increase.  We suspect
that our under sampled measurements do not clearly represent the
temporal behavior of the core luminosity.  Note that the ROSAT
luminosities are substantially higher than the average of the two EO
observations.

\subsubsection{Ambient medium}

If 26W20 is a TRG, there should be an ambient gas through which the
galaxy is moving.  Upper limits of the particle density from a hot,
X-ray emitting gas were calculated using the ROSAT HRI data.  The
particle densities estimated from different methods are displayed in
Table 4.

The particle density measurements of the sub-cluster environment
surrounding 26W20 include an acceptable range of values in agreement
with the radio non-thermal pressure measurements.  From the King model
of the cluster, it is evident that particle densities in this region
are attributed to the local environment of the sub-cluster and not to
the main cluster, Abell 754.  We can then compare our estimated upper
limit of 0.76$\times10^{-3}$ cm$^{-3}$ to that for TRGs in other clusters.  
Our upper limit falls within the range of ambient densities (0.02 -
6)$\times10^{-3}$ cm$^{-3}$ as reported by Feretti {\it et al.} (1992)
near TRGs in Abell clusters.

As a final check to detect the presence of any hot, extended gas in
the sub-cluster, 26W20 and an adjacent source were subtracted from the
HRI map.  Then the image was rebinned to $8\arcsec$ pixels and smoothed
with a Gaussian of FWHM=60\arcsec.  There appeared to be no conclusive
evidence for the existence of any hot extended gas.

\scriptsize
\begin{flushleft}
\begin{tabular}{llll}                              
\multicolumn{4}{c}{Table 4} \\
\multicolumn{4}{c}{Ambient Density Measurements} \\
\\
\bf Method&\bf Pressure&\bf Density&\bf Flux \\
&\ ($10^{-12}$ cgs)& ($10^{-3}$ cm$^{-3}$)&($10^{-13}$ cgs)\ \\ \hline
Extrapolation of King    &$< 0.6$&  $< 0.05$ \\
Model for the Main\\
Cluster$^{\it 1}$ \\
\\
Upper limits from HRI$^{\it 2}$ \\
150\("\) Radius Aperture \\
T=1 keV      & \(\leq 2.0\)& \(\leq 0.66 \)&\(\leq 2.4 \) \\
T=5 keV      & \(\leq 11.6\)& \(\leq 0.76\)&\(\leq 3.6 \)\\
\\
50\("\) Radius Aperture \\
T=1 keV      & \(\leq 4.75\)& \(\leq 1.56\)&\(\leq 0.5 \)\\
T=5 keV      & \(\leq 26.6\)& \(\leq 1.75\)&\(\leq 0.7 \) \\
\\
Radio non-thermal \\
pressure$^{\it 3}$ \\
T=1 keV      & \(\geq 0.8\)& \(\geq 0.25\) \\
T=5 keV      & & \(\geq 0.05\)  \\
\label{density}
\end{tabular}
\end{flushleft}

Notes:\\
The cgs unit of flux used is (ergs cm$^{-2}$ s$^{-1}$) and pressure is
(dyne cm$^{-2}$).  The energy range is (0.5 - 3.0 keV).\\
1) A King model for A754 with kT=7 keV (Abramopoulos and Ku; 1983)
provides an upper limit to the cluster gas. Due to an unknown
projection, 26W20 could be significantly further from the cluster
center than the projected separation.

2) Fluxes were measured for two different size regions surrounding
26W20.  The larger volume represents the scale where the radio structure
changes from straight jet/tail to a lobe. It is also the scale size
suggested by the galaxy surface density.  The smaller $50\arcsec$
region would be appropriate for the core radius if the local gas had a
King distribution.  The values are 3 sigma upper limits for a uniform
gas.

3) Since non-thermal pressure within the radio jet and lobe should be in
balance with the external medium, particularly for the lobe, a lower
limit to the pressure was calculated from the radio emission. This is
a strict lower limit since a) we assumed a filling factor of 1, b) we
integrated electron energies which correspond to only the observed
part of the radio spectrum and c) we assumed no contribution to the
energy density from relativistic protons.

\normalsize

\subsection{Optical data}

\begin{figure}
\rotatebox{270}{\resizebox{8cm}{!}{\includegraphics{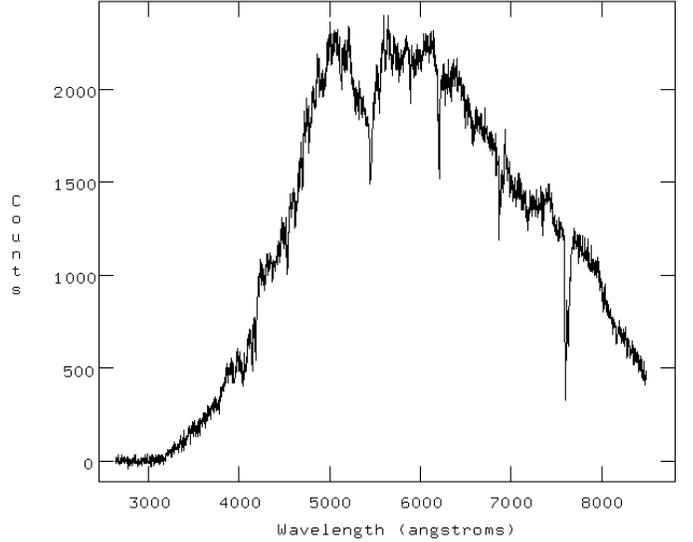}}}
\caption{Optical spectrum of 26W20 with the MMT.  Features include the
MgI absorption line at 5468 $\AA$, the CaII (H-K) break at 4162 $\AA$ 
and the NaD line at 6205 $\AA$.  Residuals from sky subtraction are
seen at about 6900 and 7600 $\AA$.  None of the usual AGN emission 
lines are found.}
\label{opt_spectrum}
\end{figure}

Two, ten minute, blue channel CCD optical spectra were taken on a
Multiple Mirror Telescope (MMT) open night on April 9, 1997.  Figure
\ref{opt_spectrum} is the summed wavelength calibrated spectrum for
both exposures.  The measured redshift of z=0.0537 agrees with
previous measurements (HCD).  The slit has a width of $1.5\arcsec$ and a
length of $190\arcsec$.  The dispersion of the grating is 300 lines/mm.
The 3K$\times$1K CCD has a spatial resolution of 0.3\arcsec/pixel.
The PA of the slit is -37.9$\degr$ aligned along the radio jet.

\section{Discussion}

The radio and X-ray core emission in conjunction with the lack of
prominent optical emission lines are properties similar to a BL Lac
type object but the lack of strong optical core emission distinguishes
this galaxy from a typical BL Lac.

Our high resolution MERLIN observation at 6 cm (Figure \ref{6cm})
shows a compact region of about 100 parsecs. These results further
support the compact nature of the core emission of the VLA images by
Harris {\it et al.} (1984) and the archival 3.6 cm data (Figure
\ref{4cm}).  The position angle of the second component (PA=140\degr)
agrees, within the errors, with those in the 21 cm map (PA=145\degr) and the
3.6 cm image (PA=150\degr).

We have computed the values for $\alpha_{ro}$(0.43) and
$\alpha_{ox}$(0.99) with an assumption that the upper limit of the
optical core flux is 0.24 mJy (1/10 of the total flux of the galaxy in
the V band).  The X-ray flux from an HRI measurement is 0.56
$\mu$Jy at 1 keV.  Given these values, 26W20 lies in the center of the
X-ray selected BL Lac objects on the $\alpha_{ro}$ vs. $\alpha_{ox}$
plot as shown in Figure 8 of Tananbaum {\it et al.} 1997.  If the core
luminosity is significantly less than our upper limit, this could
place 26W20 at an extreme position outside of the BL Lac group.  

Given the range of $\alpha_{ro}$ and $\alpha_{ox}$ for a typical X-ray
selected BL Lac, the core optical flux is between 3.2$\times10^{-2}$
and 1.55 mJy.  Therefore, an excess of optical flux in the core will
not significantly change the color or magnitude of the galaxy as a
whole.

No evidence for the presence of any hot, extended gas was found in the
sub-cluster containing 26W20.  From density measurements of the
sub-cluster environment, the values fall within the range
of ambient densities of other clusters harboring TRGs.

We propose that the BL Lac properties observed in 26W20 are evidence
of energetic emission from the core due to beamed emission from a
component of the jet.  Therefore, the large apparent (i.e. if emission
is isotropic) X-ray luminosity of $10^{43}$ ergs s$^{-1}$, concomitant with
the absence of emission lines, means either there is no gas in the
galaxy to form emission line regions or the X-rays are beamed and not
available to ionize the gas.

\begin{acknowledgements}

We thank J. Kriesslar and T. Beers for providing the galaxy number
density plot of Abell 754 and Jonathan Schachter for useful
discussions. Also, we are indebted to J. Huchra, P. Berlind, J. Mader,
S. Tokarz and D. Fabricant for the help in obtaining and analyzing an
optical spectrum of the galaxy based on observations made with the
Multiple Mirror Telescope, a joint facility of the Smithsonian
Institution and the University of Arizona.

MERLIN is a UK national facility, operated by the University of
Manchester on behalf of the  Particle Physics and Astronomy Research
Council.  We thank S. Garrington for calibrating our data.

This work was partially supported by NASA contract NAS5-30934.  An
anonymous referee contributed useful comments.

\end{acknowledgements}

\end{document}